\documentclass[aps,preprint]{revtex4}%
\usepackage{amsfonts}
\usepackage{amsmath}
\usepackage{amssymb}
\usepackage{graphicx}%
\providecommand{\U}[1]{\protect\rule{.1in}{.1in}}
\begin{document}
\preprint{ }
\title[Short title for running header]{The Legacy of M. P. Bronstein: on relativistic wave equations for spin 2
fields and some comments}
\author{Diego Julio Cirilo-Lombardo}
\affiliation{Bogoliubov Laboratory of Theoretical Physics,Joint Institute for Nuclear
Research 141980, Dubna(Moscow Region), Russian Federation}
\author{}
\affiliation{}
\author{}
\affiliation{}
\pacs{PACS number}

\begin{abstract}
We briefly give a very simple picture about one of the most remarkable results
of Matvej Petrovich Bronstein concerning the quantization of the gravitational
waves showing also that the linearized Einstein equations of the paper:
Phys.Rev. D65 (2002) 104005 are the same Bronstein's equations given 66 years before.

\end{abstract}
\volumeyear{year}
\volumenumber{number}
\issuenumber{number}
\eid{identifier}
\date[Date text]{date}
\received[Received text]{date}

\revised[Revised text]{date}

\accepted[Accepted text]{date}

\published[Published text]{date}

\startpage{101}
\endpage{102}
\maketitle
\tableofcontents

\section{M.P.Bronstein, Linearized Einstein equations and quantization of
gravitational waves}

As it is well known, the brilliant Soviet physicist M.P. Bronstein had a wide
interest in astrophysics, cosmology, semiconductors and nuclear physics but
was particularly attracted to gravitation[1,5]. Without entering too many
details (we refer to [1,5] and references therein) one of the most significant
Bronstein's works " The quantization of the gravitational waves" was published
in 1936 [2], being the first deep investigation in quantum gravity. The main
goal of his paper [2] was the presentation of the Einstein equations in the
following form:%
\begin{align}
\frac{1}{c}\frac{\partial H_{ij}}{\partial t} &  =-\varepsilon_{ikl}%
\frac{\partial E_{lj}}{\partial x_{k}},\text{ \ \ \ \ \ \ }\frac{\partial
H_{ij}}{\partial x_{i}}=0\tag{1}\\
\frac{1}{c}\frac{\partial E_{ij}}{\partial t} &  =\varepsilon_{ikl}%
\frac{\partial H_{lj}}{\partial x_{k}},\text{ \ \ \ \ \ \ }\frac{\partial
E_{ij}}{\partial x_{i}}=0\tag{2}%
\end{align}
where $E_{ij}$ and $H_{lj}$ are symmetrical and traceless tensors defined as
(in Bronstein's work original notation,$\ x_{0}=it$)
\begin{align}
E_{ij} &  =R_{4i4j}=\frac{1}{4}\varepsilon_{ikl}\varepsilon_{jmn}%
R_{klmn}\tag{3}\\
H_{ij} &  =\frac{i}{2}\varepsilon_{imn}R_{mn4j}=\frac{i}{2}\varepsilon
_{imn}R_{4jmn}\tag{4}%
\end{align}
and $R_{klmn}$ is the curvature tensor (notice that this fact is not
accidental, it follows from the general form of the relativistic wave
equations for massless fields [4])

More recently in the paper [3] by E.T. Newman the following equations appear
(in Newman's work notation)%
\begin{equation}
\nabla_{a}W^{abcd}=0\tag{5}%
\end{equation}
where in the above equations the following selfdual quantity was defined
\begin{equation}
W^{abcd}=C^{abcd}-iC^{\ast abcd}\tag{6}%
\end{equation}
where $C^{abcd}$ is the Weyl tensor, that as it is well known, it coincides
with the Riemmann tensor in the case of Einstein equations in vacuum without
cosmological constant. Consequently eq.(5) are nothing more than, namely, the
second Bianchi identities. Then, these are exactly the Bronstein equations
(1,2) because after the 3+1 splitting the complex field equations (5) take the
form
\begin{align}
\nabla_{0}W^{0i0k}+i\varepsilon^{ijl}\nabla_{j}W^{0l0k} &  =0\tag{7}\\
\nabla_{i}W^{i0j0} &  =0\tag{8}%
\end{align}
where now if we redefine $W^{0i0k}$ as
\begin{equation}
W^{i0j0}\equiv E^{ij}+iH^{ij}\tag{9}%
\end{equation}
(in complete analogy within the case of the Maxwell equation in flat space) we
immediately obtain
\begin{align}
\nabla_{0}\left(  E^{ik}+iH^{ik}\right)  +i\varepsilon^{ijl}\nabla_{j}\left(
E^{lk}+iH^{lk}\right)   &  =0\tag{10}\\
\nabla_{i}\left(  E^{ij}+iH^{ij}\right)   &  =0\tag{11}%
\end{align}
that are nothing more than the Bronstein equations:%
\begin{equation}
\nabla_{0}E^{ik}=\varepsilon^{ijl}\nabla_{j}H^{lk},\text{ \ \ \ \ \ \ \ }%
\nabla_{i}E^{ij}=0\tag{12}%
\end{equation}%
\begin{equation}
\nabla_{0}H^{ik}=-\varepsilon^{ijl}\nabla_{j}E^{lk},\text{ \ \ \ \ }\nabla
_{i}H^{ij}=0\tag{13}%
\end{equation}
\ (considering that linealization means flat metric connection, as pointed out
also in [3]). However and surprisingly into the paper [3] ( see Section III:
Linealized GR ) no mention about the Bronstein work nor his specific reference
[2] were given. Only, the author of [3] redefines again the expression (9) as
$E^{ij}+iH^{ij}\equiv Z^{ij}$ rewritting equation (7) and (8) (alternatively
(10) and (11)) as%
\begin{align}
\partial_{t}\overleftrightarrow{Z}+i\cdot\operatorname{curl}%
\overleftrightarrow{Z} &  =0\tag{14}\\
\operatorname{div}\overleftrightarrow{Z} &  =0\tag{15}%
\end{align}
where $\overleftrightarrow{Z}=\overleftrightarrow{E}+i\overleftrightarrow{H}$
was defined in [3] as "dyadic form". The equations were concluded in [3] with
the following claim (the numbers of the formulas correspond to our paper):

" \textbf{Remark 1} \textit{It has been well known for many years that Eq.(5)
can be viewed as the linear Einstein equations. We, however, have not been
able to find in the literature the particular form, (14,15), that so mimics
the Maxwell equations. Though we would be surprised, it might well be new."}

Sometimes certain phrases do not ever lose value despite their age, social or
historical context depending only on the wisdom of the speaker:

"\textit{What was, will be again, what has been done, will be done again, and
there is nothing new under the sun! Take anything which people acclaim as
being new: it existed in the centuries preceding us. No memory remains of the
past, and so it will be for the centuries to come -- they will not be
remembered by their successors}." King Solomon in Ecclesiastes 1, 9-11

As always, the greatness of a person is in his capacity to recognize their
failures, successes and mistakes. A possessor of this greatness (in my humble
opinion) is without any doubt the referee of this paper and author of
reference [3]. He himself has asked me to add a comment, and I have the
pleasure to do it:

"\textbf{Remark 2 }\textit{I am delighted to have it pointed out that the work
of the brilliant Russian physicist M. P. Bronstein preceded my work by 66
years}."

"\textbf{Remark 3} \textit{The author of reference [3] kindly inform us about
that in ref.[6] there is the construction of the gravitational wave equations
given by M. Bronstein but no mention about ref. [2]}"

\section{Acknowlegdements}

We are very grateful to the people of the Bogoliubov Laboratory of Theoretical
Physics (BLTP) and JINR Directorate by they hospitality and financial support,
and also to Professor Yu. P. Stepanovsky who introduced me to the subject of
relativistic wave equations and to the life and research of M. P. Bronstein.
Special thanks are given to Professor Stanley Deser and to Professor Gennadi
Gorelik for bring us useful suggestions and main references to the subject and
particularly to Professor Ezra T. Newman that kindly reviewed this manuscript.

\section{References}

[1] G. E Gorelik and V.Ya. Frenkel,\textit{ Matvej Petrovich Bronstein
}(1906-1938), Nauka(Moscow) 1990.

[2] M.P. Bronstein, Zh.ETF (JETP), V6 (1936) 195 (in russian).

[3] E.T.Newman, Phys.Rev. \textbf{D65} (2002) 104005.

[4] Yu.P. Stepanovsky, Ukr.Fiz.Zh.(Ukr.Ed.) 9 (1964) 1165-1168

[5] S. Deser and A. Starobinsky, Introduction to translation of Bronstein's
original 1935 paper as a "Golden oldie", Gen. Rel. Grav. 44, 263-265 (2012)

[6] Roy Maartens and Bruce A. Bassett, Class. Quantum Grav. 15, 705, 1998.

\end{document}